\documentclass[preprint2]{aastex}

\slugcomment{Astronomical Journal}

\shorttitle{Discovery of New Nearby Stars}
\shortauthors{Hambly et al.}

\begin{document}

\title{The Solar Neighborhood VIII: Discovery of New High Proper
Motion Nearby Stars Using the SuperCOSMOS Sky Survey}

\author{Nigel C.\ Hambly}

\affil{Institute for Astronomy, School of Physics, University of Edinburgh,}
\affil{Royal Observatory, Blackford Hill, Edinburgh, EH9~3HJ, Scotland, UK}

\author{Todd Henry, John Subasavage, Misty Brown and Wei--Chun Jao}

\affil{Department of Physics \& Astronomy}
\affil{Georgia State University, Atlanta, GA 30303--3083}


\begin{abstract}

Five new objects with proper motions between 1.0\arcsec/yr and
2.6\arcsec/yr have been discovered via a new RECONS search for high
proper motion stars utilizing the SuperCOSMOS Sky Survey.  The first
portion of the search, discussed here, is centered on the south
celestial pole and covers declinations $-$90$^\circ$ to
$-$57.5$^\circ$.

Photographic photometry from SuperCOSMOS and $JHKs$ near-infrared
photometry from 2MASS for stars nearer than 10 pc are combined to
provide a suite of new M$_{Ks}$-color relations useful for estimating
distances to main sequence stars.  These relations are then used to
derive distances to the new proper motion objects as well as
previously known stars with $\mu$ $\ge$ 1.0\arcsec/yr (many of which
have no trigonometric parallaxes) recovered during this phase of the
survey.

Four of the five new stars have red dwarf colors, while one is a
nearby white dwarf.  Two of the red dwarfs are likely to be within the
RECONS 10 pc sample, and the white dwarf probably lies between 15 and
25 pc.  Among the 23 known stars recovered during the search, there
are three additional candidates for the RECONS sample that have no
trigonometric parallaxes.

\end{abstract}


\section{Introduction}

One of the primary goals of the Research Consortium on Nearby Stars
(RECONS) is to discover stars that are currently missing from
compendia of the Sun's nearest neighbors.  As discussed in
\citet{hen97}, we estimate that as many as 30\% of the nearby star
systems lurk undiscovered within 10 parsecs, the horizon of the RECONS
sample.  The RECONS team is using both of the methods typically used
to reveal nearby stars --- high proper motion and photometric distance
estimates --- to discover new nearby systems.  Here we report first
results of a new initiative resulting in the discovery of five new
high proper motion objects with $\mu$ $>$ 1.0\arcsec/yr in the region
around the south celestial pole, along with an analysis of the newly
revised sample of all similarly defined high proper motion stars in
the same region.

To date, Luyten and Giclas have provided the bulk of high proper
motion star systems.  The valuable {\it LHS Catalogue} 
\citep[hereafter LHS]{luy79a}
is a compendium of stars that includes 3583 objects
with annual proper motions greater than 0.5\arcsec/yr.  Included in this
list are 525 stellar systems that have motions greater than
1.0\arcsec/yr.  Many of these systems were also reported by Giclas and
collaborators \citep{gic71}.  Both the Luyten and Giclas surveys were
monumental undertakings in an era before extensive computer power was
available and electronic data mining possible. Moreover, not all parts
of the sky were covered uniformly in those surveys.  Significantly
lower completeness in the most southerly declinations is particularly
noteworthy.

There is no doubt as to the scientific usefulness of the monumental
Luyten and Giclas surveys; however there are two key shortcomings that
have always dogged efforts to exploit those catalogues for
quantifiably complete samples.  First, the absolute astrometric
accuracy of Luyten's catalogue is poor by modern standards, making
recovery of some of the catalogued objects very difficult
\citep[eg.][]{bak02}.  
This is despite the availability of finder charts
in the {\it LHS Atlas} \citep{luy79b}.  Second, and more
significantly, the completeness of surveys produced using a variety of
techniques (including manual plate `blinking') has been difficult to
establish.  Worries concerning the completeness of the LHS have been
compounded over the past decade by a spate of new discoveries, for
many of which there is no obvious reason as to why they were missed in
the earlier surveys.  Adding further fuel to the
firey completeness debates is the fact that invariably newly
discovered objects turn out to be rather interesting, because
completeness is chiefly an issue at relatively faint magnitudes and
high proper motions where new types of intrinsically faint, high
velocity and potentially nearby stars tend to be found
\citep[eg.][]{ham97}.  Clearly, if there is a large and unquantifiable
bias against certain classes of objects in the early surveys, then
statistical corrections \cite[eg.][]{daw86} can never hope to recover
an accurate estimate of the true numbers of all types of star.
Recently, however, the availability of homogeneous, multi--colour and
multi--epoch Schmidt survey plate collections along with fast, high
precision scanning machines capable of digitising them has enabled
significant progress.  An example is the SuperCOSMOS Sky Survey
(hereafter SSS, \citealt{ham01a}; 
that paper also briefly describes some
of the other major digitisation programmes).  These newly digitised
sky surveys have enabled much progress in systematic trawls for high
proper motion stars \citep[eg.][]{iba00}.  Completeness questions can
now be more accurately addressed \citep[eg.][]{mon00} and recovery of
many of Luyten's discoveries has become possible \citep[eg.][]{bak02},
in addition to discovery of new objects.

Thus, recent astrometric efforts to reveal high proper motion stars
are meeting with continued success.  Several studies investigating the
southern sky have yielded new systems with $\mu$ $>$ 1.0\arcsec/yr.
The large effort of Wroblewski and collaborators (\citealt{wro01} and
references therein) has resulted in discovery of two systems (WT 248
and WT 1827).  The Calan-ESO survey of Ruiz and collaborators
(\citealt{rui87} and \citealt{rui01}) has yielded three new systems (ER2,
ER8, and CE 89).  Work by Scholz and collaborators has used APM
measurements of UK Schmidt Telescope survey plates (eg.~\citealt{rey02};
\citealt{lodi04}; \citealt{sch00}) 
and the SSS (eg.~\citealt{sch02} and references therein)
to reveal three systems (APMPM J1957--4216, APMPM J2330--4736, and
SSSPM J2231--7514AB).  The study of \citet{opp01} also used the SSS,
and found new white dwarfs (eg.\ WD0205--053, WD0351--564, and
WD2214--390).  Furthermore, \citet{pok03} report a systematic high
proper motion star survey employing SSS data; detailed follow-up of
potentially interesting objects is underway by our group and others.

Astrometric searches of the northern sky also continue to yield high
proper motion stars.  In an innovative, systematic search of the sky
north of declination $-$2.8$^\circ$ and within 25$^\circ$ of the
Galactic plane, \citet{lep02} have found 18 new systems with $\mu$ $>$
1.00\arcsec/yr, by far the most productive recent effort, while
\citet{tee03} have reported a remarkable system with $\mu$ $=$
5.06\arcsec/yr by searching the SkyMorph database of Near--Earth
Asteroid tracking data \citep{pra99}.

In an effort to discover previously missed high proper motion objects
in the southern sky, a new search by the RECONS team was initiated
using SSS data.  The new objects have been dubbed SCR sources,
corresponding to the SuperCOSMOS RECONS search.  In this paper we
present the first results of this search, which includes five new
objects with motions larger than 1.0\arcsec/yr in only $\sim8$\% of
the entire sky.  We have extracted $JHKs$ photometry from 2MASS to provide an
extended color baseline that permits accurate photometric distance
estimates.  In a follow--up paper \citep{hen04} we will present
further optical photoelectric photometry, spectroscopy and analysis of
accurate photometric parallaxes for the sample.

\section {Trawling SuperCOSMOS Sky Survey data}

The SSS \citep{ham01a} is ideally suited to systematic searches for
high proper motion stars.  The survey consists of multi--colour $(BRI)$
Schmidt photographic observations with $R$ at two distinct epochs, but
in general all four passbands (plate magnitudes denoted by $B_{\rm
J}$, $ESO-R$, $R_{\rm 59F}$, and $I_{\rm IVN}$, and hereafter referred
to as $B_{\rm J}$, $R_1$, $R_2$ and $I$) were observed at different
epochs separated by up to 50 years.  Data for the entire southern
hemisphere are currently publicly available (see \citealt{ham01a} for
details) and the survey programme is currently being extended into the
northern hemisphere.  As described above, there have been several
searches for high proper motion stars using SSS data.  Each has had a
different emphasis, with different science goals.

The survey of \citet{pok03} is an automated search primarily using the
two $R$ band plates (UK and ESO Schmidt surveys) in each field of the
SSS south of $\delta$ $=$ $-$20$^\circ$, aimed at cataloguing cool
dwarfs to a lower proper motion limit of $\mu$ $=$ 0.18\arcsec/yr.  The
search methodology employs software--automated multiple--pass pairing
between only two $R$ plates, and necessarily has rather stringent limits
on image quality (eg.~profile class, proximity to bright stars and
general morphology) to yield a clean catalogue of thousands of stars;
the intention was that no manual sifting of this automatically generated
catalogue would be required.  The chief
limitations of the search method are general incompleteness resulting
from the stringent quality constraints, restriction of sky area
covered due to necessary avoidance of crowded regions and
incompleteness at high proper motions -- c.f.~\citet{pok03} Section~4
-- due to spurious pairing at large image displacement.

The trawl employed by \citet{opp01} was similar to that of Pokorny et
al.\ but was aimed at cool (but relatively blue) white dwarfs.
Detections therefore relied on presence on the $B_{\rm J}$ plates as
well as the two $R$ plates, and hence less stringent quality cuts
could be used to obtain a clean sample.  This survey was also
generalised to use $POSS-I$ `E' ($\equiv$ $R$) data in the equatorial
zone $0^\circ > {\rm Dec} > -20^\circ$ in lieu of $ESO-R$ material,
but employed a proper motion threshold of $\mu$ $=$ 0.33\arcsec/yr.
In this case, some visual checking of the available 
source image data was employed after
selection of stars via colour and reduced proper motion.  However, the
\citet{opp01} survey also suffered from the same incompleteness
problems at high proper motions, and inability to probe more crowded,
lower latitude regions of sky.

\subsection {New search methodology}

For the purposes of supplying candidates for RECONS, a new search
strategy is being employed that attempts to circumvent the
completeness problems of previous SSS efforts by a combination of full
use of all astrometric information between the four plates available
in every field, and relaxed quality/morphology criteria along with a
final step of manual sifting to produce clean target lists.  In some
ways this is an unsatisfactory return to the Luyten--style approach by
including a subjective human element to the process, but it seems that
at least for surveys employing parameterised image detection lists
this is the only way of ensuring high levels of completeness.
Ultimately, it is likely that a whole--sky application of a
pixel--based process like the SUPERBLINK algorithm of \citet{lep02} is
the only way of attaining the highest possible completeness from
digitised Schmidt plate scans; even then the problems of saturation
and scattered light near bright stellar cores remain (hence, areal
completeness and completeness for close binaries will never be close
to 100\%).

Briefly, the new SCR search starts with each individual set of
parameterised detections (the so--called Image Analysis Mode data --
see \citealt{ham01b}) from all plates in the same field positionally
error--mapped to a common co-ordinate system using a rigorous
application of the \citet{eva95} error mapping algorithm.  The default
SSS pairing \citep{ham01c} is then used to exclude all images that
appear on all four plates having astrometric solutions indicating
proper motion less than the limit (here, 0.4\arcsec/yr) along with
goodness--of--fit parameter $\chi^2<1.0$.  Then, all images that are
either unpaired or have inconsistent astrometric solutions (resulting
from erroneous matching in the simple default SSS scheme) are
processed one by one in all possible combinations amongst the
available measurements out to a maximum displacement dictated by the
upper proper motion limit chosen for the survey (here, 10\arcsec/yr)
and the maximum epoch difference between the set of plates.  This
`brute force' approach is made possible by modern computers and
storage which are capable of processing large amounts of data at high
IO bandwidth.  The other innovation in our latest SSS trawl for RECONS
is that single cuts in a range of quality/morphology parameters are
not made; rather a set of warning conditions is defined at levels set
by maximising completeness with respect to the LHS (at the same time
endeavouring to minimise contamination by spurious detections to make
the final manual sifting problem tractable).  A violation threshold is
set such that only when three quality conditions are violated for a
four plate detection (two for a three plate detection) will a
candidate be thrown away as certainly dubious.

\begin{figure*}
\epsscale{2.0}
\plotone{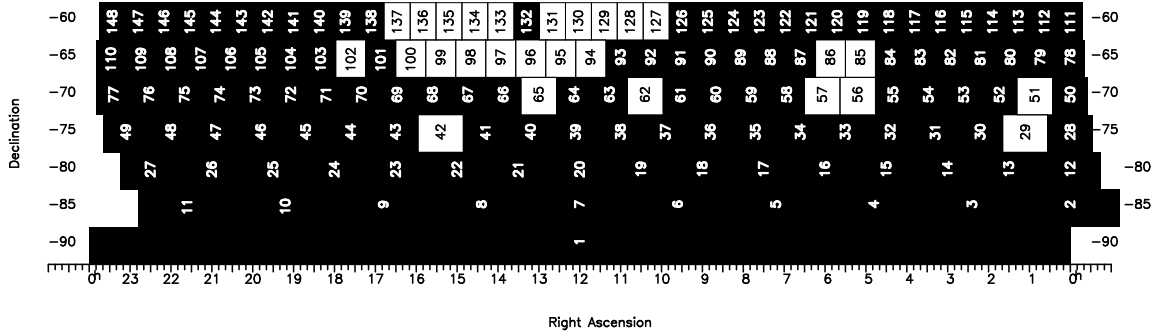}
\caption{ESO/SRC standard fields covered in the SCR survey so far.
Black cells indicate included fields; white cells indicate that the
field was too crowded to be reliably analysed using the current data
and algorithm (such fields have an unfavourable spread of epochs
amongst the four SSS plates or are heavily crowded, being at low
Galactic latitudes or near the cores of the Magellanic Clouds).  Each
field covers $\sim25$ square degrees of distinct sky (ie.~not
including any overlap regions) and we have covered 121 out of 148
possible ESO/SRC fields.  Hence, the sky area covered so far is
$\sim3000$ square degress, or $\sim8$\% of the whole celestial
sphere.}
\label{fields}
\end{figure*}

The SCR search is being carried out starting at the south celestial
pole and is moving north.  This paper describes the first phase of the
search, which includes the region from $\delta$ $=$ $-90^\circ$ to
$-57.5^\circ$.  Figure~\ref{fields} shows the fields currently
surveyed.  Number labels in Figure~\ref{fields} are ESO/SRC standard
Schmidt survey field designations; black areas represent fields
included while white areas indicate fields currently excluded due to
crowding problems (areas of low Galactic latitude or areas within the
Magellanic Clouds).  In total, $\sim8$\% of the sky has been searched
in this first effort.

The initial sift of the SSS dataset for RECONS included sources with
proper motions determined to fall between 0.4\arcsec/yr and
10.0\arcsec/yr, and magnitude ${\rm R}<16.5$.  The primary goal is to
discover objects moving faster than $\mu$ $=$ 0.5\arcsec/yr, but the
lower boundary on $\mu$ was chosen to be 0.4\arcsec/yr so that LHS
objects with motions near 0.5\arcsec/yr could be recovered if computed
proper motion was slightly less than the cutoff due to measurement
errors in either SSS data or the LHS.

\subsection {Statistics of Results}

The total number of candidate objects detected with $\mu$$_{SCR}$ $=$
0.4--10.0\arcsec/yr in this initial search was 897.  Spurious high
proper motion detections can arise due to plate
defects, blended sources and halo `sources' around bright stars (for
example, we note that the list of~\citealt{pok03} 
contains 8 apparently spurious
sources having $\mu > 4$\arcsec/yr -- one at more than 10\arcsec/yr).
A multi-step sifting process was used to vet the SCR search candidates for
true and false detections, including checks of magnitudes, colors, and
image ellipticities.  The $R_1$ and $R_2$ magnitudes were checked for
consistency, and the colors were examined to determine whether they
matched that of a real object, i.e. both $B-R_2$ and $B-I$ positive, or
both negative.  If the candidate passed both checks, it was passed on to 
the visual inspection stage.

In cases where a candidate failed the first two tests, the ellipticity
quality flag was also checked.  Experience revealed that if two or more
image ellipticities were larger than 0.2, the object was spurious.  
Detections that failed all three tests were classified as false without
visual inspection.  As a final check, all of the 99 candidates found
between $\delta$ $=$ $-$90 and $-$80 were inspected visually (regardless
of the checks), and all fell into the appropriate true or false detection
bins.

For the true detections, coordinates were cross-correlated with the SIMBAD
database and the NLTT catalog.  If the coordinates agreed to within a few
arcminutes and the magnitudes and proper motions were consistent, the
detection was labeled as previously known.  For detections without known
proper motion counterparts, visual inspection definitively confirmed or
refuted a real proper motion object.  In a few cases, the coordinates
agreed well, but the magnitudes did not.  Several of these near matches
turned out to be new common proper motion companions to a previously known
proper motion object.

In summary, 
these checks revealed 443 false detections (49\% of the candidates);
additionally, 72 of the 897 candidates were duplicate detections
resulting from the generous ESO/SRC plate overlap regions.  Of the
resulting 382 real, distinct objects, 262 were recoveries of
previously cataloged objects, including many LHS/LTT objects and stars
from the other surveys mentioned in Section~1.  Finally, 120 (13\% of
the initial candidate list) were found to be new discoveries,
including a handful of new common proper motion companions to already
known primaries.

\section {Data from SuperCOSMOS --- Astrometry and Plate Photometry}

Of the 120 new objects discovered and the 262 known objects recovered
in this phase of the SCR effort, we concentrate here on the subsample
of 28 stars with $\mu$$_{SCR}$ $\ge$ 1.0\arcsec/yr, 
which are listed in Table 1.
Names for the five new SCR stars are given in the first column (finder
charts are given in Figure 2), whereas the known names are given for
the remaining 23 stars.  Also listed are the SSS photographic
astrometry and photometry for all 28 objects.  Coordinates are
precessed and proper motion corrected to equinox and epoch J2000.0,
and are accurate to $\pm0.3$~arcsec.  Proper motions (and their
errors) and the position angles for each target are given.

\begin{figure*}
\epsscale{2.0}
\plotone{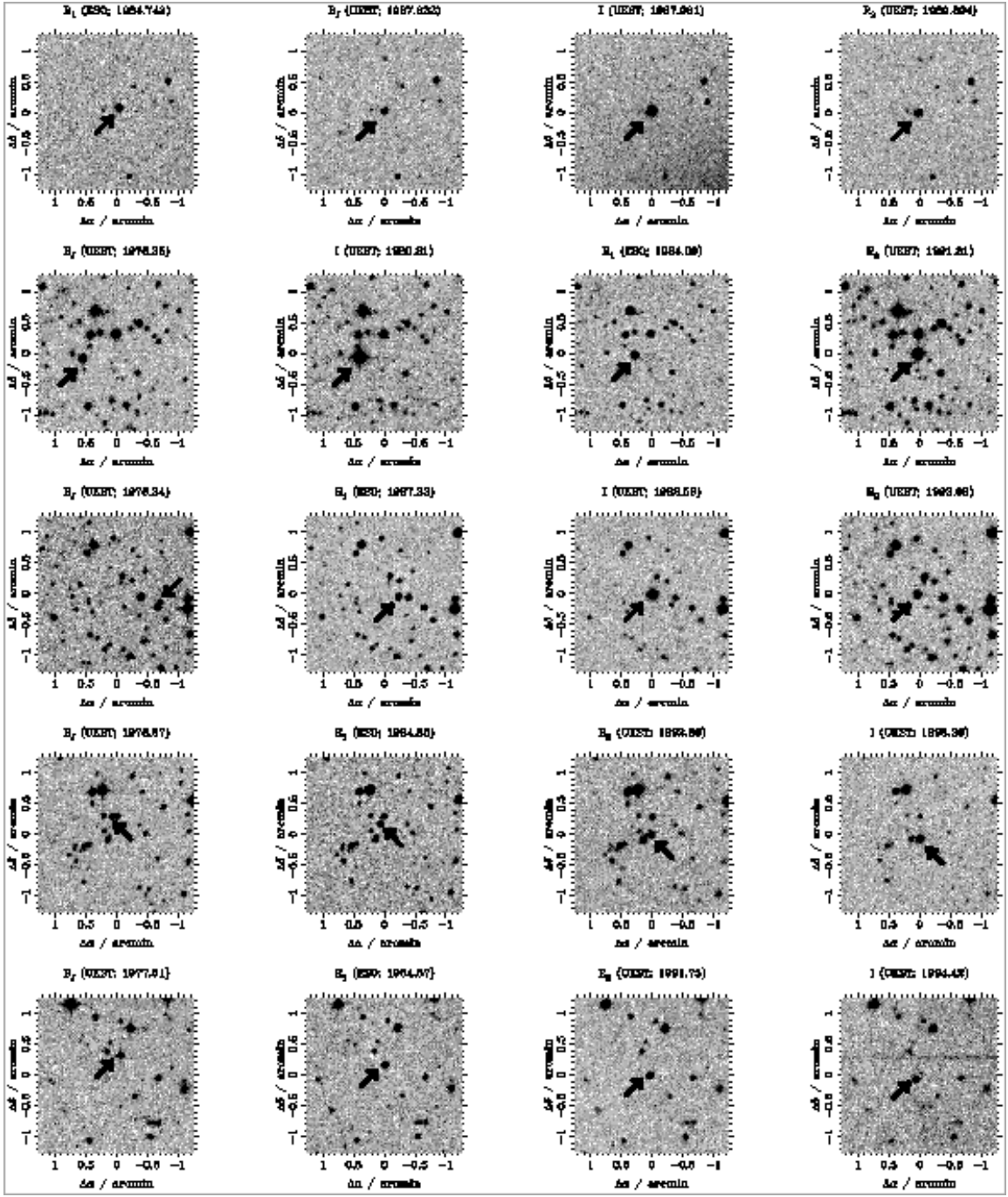}
\caption{`Postage stamp' finders for the five newly discovered high
proper motion stars.  Rows are (i) SCR0342--6407; (ii) SCR1138--7721;
(iii) SCR1845--6357; (iv) SCR1848--6855; and (v) SCR2012--5956.  In
each case, the four images are the $B_{\rm J}$, $R_1$, $R_2$ and $I$
images in chronological order.  Black arrows indicate the star in
question in each case.}
\label{finders}
\end{figure*}

We quote photometry in Table~\ref{photography} from the UK Schmidt R
original survey plates ($R_2$) because these data are, in general, of
higher signal--to--noise and more uniformly calibrated than that from
the ESO--R copy plates ($R_1$).  In addition, red objects sometimes do
not have reliable detections at $B_{\rm J}$, while blue objects are
often faint on $I$ plates.  Note that the absolute calibration of the
individual passbands is subject to systematic errors that increase as
the brightness of the source increases \citep{ham01b}.  However,
because corrections to colours are applied to the data with respect to
the $B_{\rm J}$ plates, these systematic errors are not present in
colour indices (eg.\ $B_{\rm J}-R_2$, $R_2-I$).  Consequently, the
relative accuracy of SSS colour indices is much better than the
absolute accuracy of individual passband photometry.  As expected, the
five new stars are fainter ($R_2$ $=$ 14.12 to 16.33) than nearly all
of the known stars ($R_2$ $=$ 7.49 to 14.99) except the double white
dwarf system ($R_2$ $=$ 15.82 and 16.21) found by \citet{sch02}
using SuperCOSMOS data.

\subsection {Completeness and other checks}

We make no claim as to the absolute completeness of our new SCR search
at this preliminary stage.  However, we note that in the area
surveyed, there are 169 LHS stars; we recover 127 of these.  At first
sight, this success rate of 75\% seems rather poor, but this test
requires closer examination.  The LHS catalog is biased (particularly
so in the southern hemisphere) towards brighter magnitudes, and the
SCR search employs deep, sky--limited Schmidt survey plates upon which
stars with ${\rm m}<10$ have heavily saturated and large, extended
images.  Moreover, the surveyed region in Figure~\ref{fields} is at
generally low Galactic latitude.  LHS objects missed by our procedure
are lost because of blending problems on the source plates and the
consequent failure of the standard SuperCOSMOS image analysis software
\citep{ham01b} in unscrambling and/or accurately parameterising
deblended components.  This is illustrated in Figure 2, where the
$B_{\rm J}$ images of SCR1845--6357 and SCR1848--6855 are blended to
such an extent as to render their $B_{\rm J}$ parameters unusable
(eg.\ Table~\ref{photography}).  Blending is particularly problematic
for brighter images, and as already stated the deep sky--limited
Schmidt survey plates are not well suited to studies of stars brighter
than ${\rm m}\sim10$.  If we limit the LHS completeness comparison to
magnitudes ${\rm R}>10.0$, we recover 112 out of 130 stars --- a
somewhat improved 86\% success rate.  If we further limit the
comparison to stars with $\mu$ $>$ 1.0\arcsec/yr, we recover 18 out of
18 LHS stars --- 100\% success.  Hence our search is most successful
in the region of parameter space where the LHS is least complete: at
fainter magnitudes and high proper motions.  Note that images of moving
stars are more susceptible to crowding than non--moving stars.  If a
high proper motion star is irretrievably blended on one R plate or on
two or more of any of the four SSS plates, the SCR trawl will not
detect it.  Images of slow moving stars are, of course, likely to be
isolated on all plates if they are isolated on one; this is not the
case for fast moving stars, especially when they are traversing a
crowded field.  If we compare SCR success versus LHS for $\mu$ $>$
1.0\arcsec/yr {\em without} a magnitude cut of ${\rm R}>10.0$, we
recover 20 out of 29 stars -- this low success rate of 69\%
illustrates the difficulty of finding bright, high proper motion stars
using deep, sky--limited Schmidt plates.  These 20 stars are listed in
Table~\ref{photography} with their LHS numbers.

\begin{figure*}
\plotone{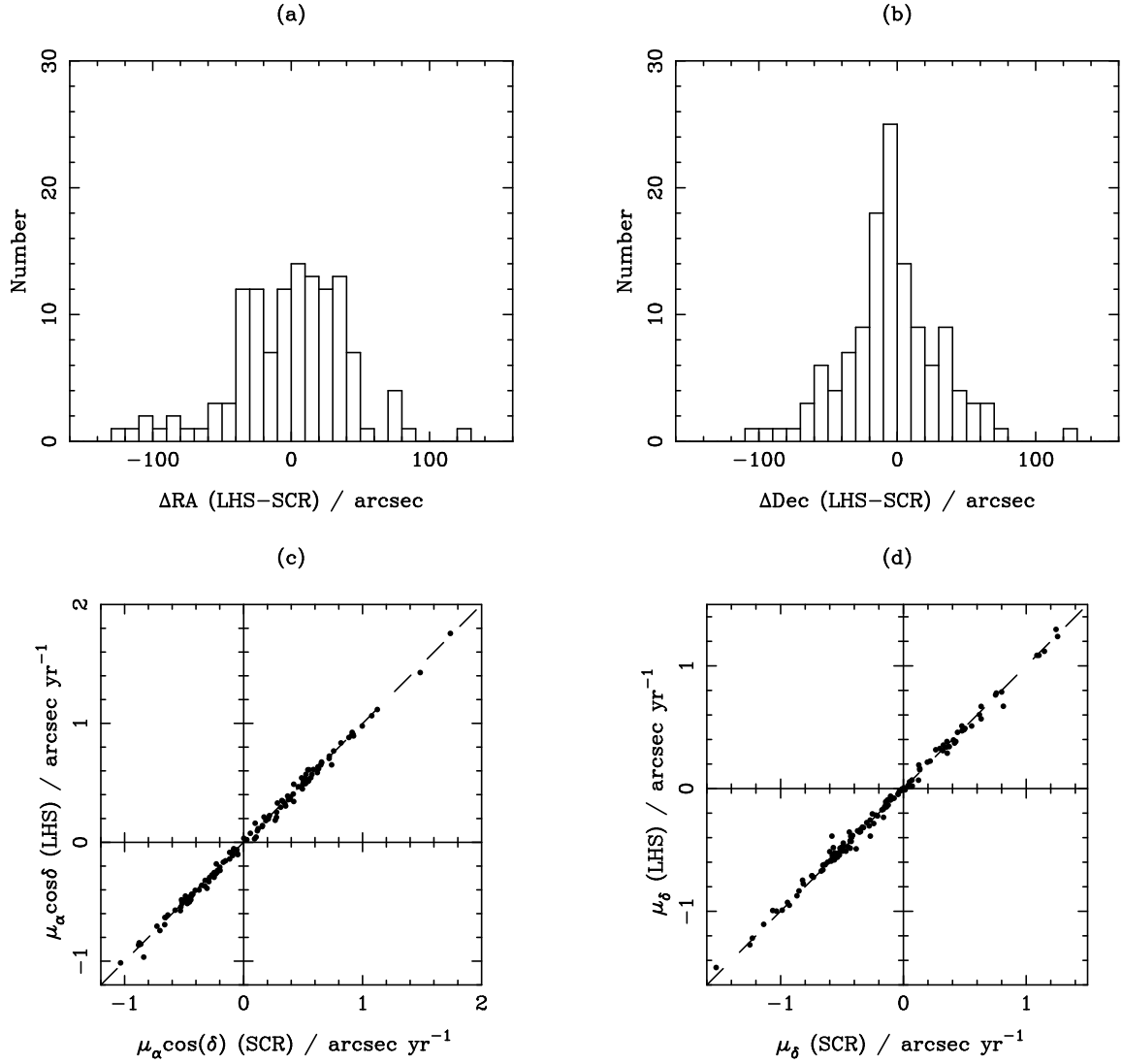}
\caption{Comparison of LHS astrometry with new data from the SCR
search: (a) Right Ascension; (b) Declination; (c) proper motion in RA;
and (d) proper motion in Dec.  Dashed lines in (c) and (d) are $y=x$
lines indicating perfect agreement.}
\label{complhs}
\end{figure*}

\begin{figure*}
\plotone{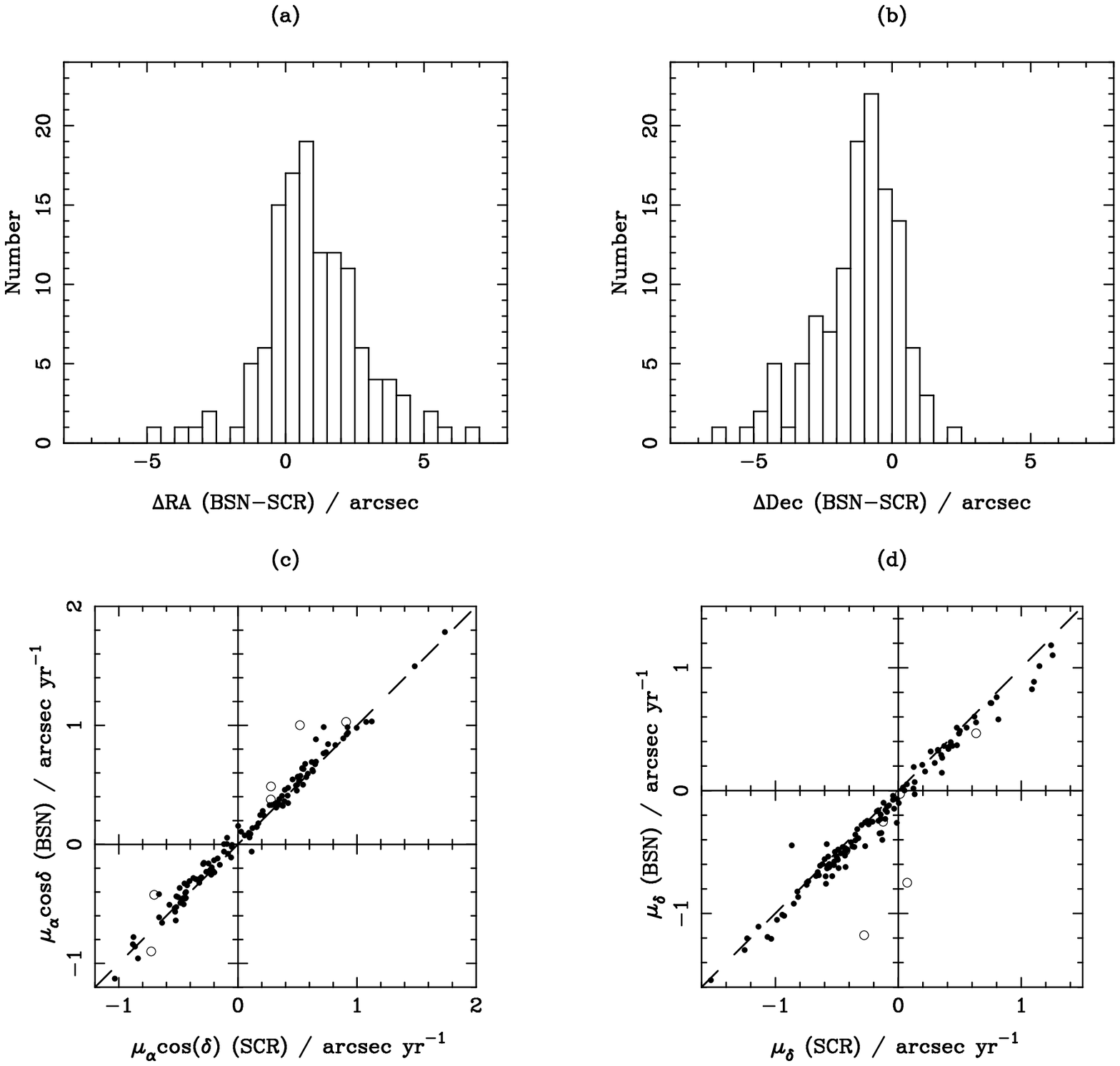}
\caption{Comparison of revised astrometry (Bakos et al.\ 2002) with
new data from the SCR search for LHS stars: (a) Right Ascension; (b)
Declination; (c) proper motion in RA; and (d) proper motion in Dec.
Open circles in (c) and (d) indicate sources for which Bakos et al.\
report unreliable proper motion measurement ``for some reason''.}
\label{compbsn}
\end{figure*}

The sample of 127 recovered LHS objects provides a control against
which the SCR astrometric measurements were compared.  In
Figures~\ref{complhs} and~\ref{compbsn} we show a comparison of our
astrometric results with those of the LHS as originally published and
using revised data \citep{bak02}.  In both cases, positions have been
precessed and proper motion corrected to a common equinox and epoch
(J2000.0 for both).  Interestingly, although the Bakos et al.\
positions are far superior to the original LHS values, their proper
motion determinations are clearly inferior, even allowing for the
objects labelled `B' or `b' in their list (open circles in
Figure~\ref{compbsn}).  The SCR and LHS proper motions are in much
better agreement, indicating the quality of Luyten's original
measurements.  While the Bakos et al.\ positions are a vast
improvement and enable easy recovery of LHS stars (eg.\ as has been
done here), their proper motion estimates should not be used in place
of the original LHS measurements.

\section {Data from 2MASS --- Infrared Photometry}

The infrared $J$, $H$, and $Ks$ photometry has been extracted from
2MASS by OASIS. Because these objects are high proper
motion stars, all of them were manually identified by comparison with
finding charts instead of retrieving data by setting a search radius 
around a given RA and Dec. 
The photometry is given in Table~\ref{photometry} for both the
new SCR stars and the known stars. The errors (and here we adopt the
$x_msigcom$ errors where {\it x} is {\it j}, {\it h}, or {\it k}) that
give a measure of the total photometric uncertainty, including global
and systematic terms) are 0.02 to~0.03 mag in most cases, and are less
than 0.05 mag in all cases except SCR2012--5956 (0.11 at $H$ and no
given error at $Ks$ because it is at the limit of $Ks$ band
detectability), J2231--7515 (0.06 at $H$ and 0.12 at $Ks$), and
J2231--7514 (0.06 at $H$ and 0.08 at $Ks$).  The 2MASS frames have also
been used to confirm the proper motion for all five newly discovered
stars.
 
The infrared photometry is useful because it permits a color extension
from the optical $B_{\rm J}$ band to $Ks$, thereby spanning more than
a factor of four in effective wavelength.  Diagnostics bridging the
photographic and infrared bands are particularly good for the
detection of blue and very red objects.  $(R_2-Ks)$ is given in
Table~\ref{photometry} as a color indicator because for the faintest
red objects $B_{\rm J}$ is not available.  Given that SCR2012--5956 is
a fast-moving, faint object, yet appears rather blue, with $(R_2-H)$
$=$ 0.40 (the value for $R_2-Ks$ $=$ 0.22 is suspect because of the
2MASS faint $Ks$ limit), it is almost certainly a white dwarf.
Computation of its reduced proper motion ${\rm H}_{\rm R} = {\rm
m}_{\rm R}+5\log\mu+5$ and ($B_{\rm J}-R_2)$ places it firmly in the
region of spectroscopically confirmed white dwarfs in Figure~1 of
\citet{opp01}.  It is also evident that SCR1845--6357 is a very red
object $(R_2-Ks$ $=$ 7.82$)$, and therefore likely to be quite close,
given it's bright apparent magnitudes.

\section {New Relations for Estimating Photometric Distances}

In order to develop reliable color-M$_{Ks}$ relations, both
SuperCOSMOS and 2MASS were searched for stars in the RECONS 10 pc
sample.  These stars are used because they generally have high quality
trigonometric parallax values and have been vetted better than any
other sample of stars for close companions that would corrupt flux and
color measurements.

Photographic $B_{\rm J}$, $R_2$ and $I$ magnitudes were extracted for
all stars south of the current declination cutoff of SuperCOSMOS
($+3.0^\circ$).  Single stars with reliable (unblended, unsaturated)
photographic magnitudes were then examined in 2MASS, from which $JHKs$
photometry was obtained.  The final cut provided 54 main sequence,
single, stars with reliable magnitudes in all six bandpasses.  These
were supplemented with one additional object, GJ 1001 B $=$ LHS 102 B,
an L dwarf found in all bands but $B_{\rm J}$ (the complete sample
and magnitudes can be obtained from the authors upon request).

\begin{figure*}
\plotone{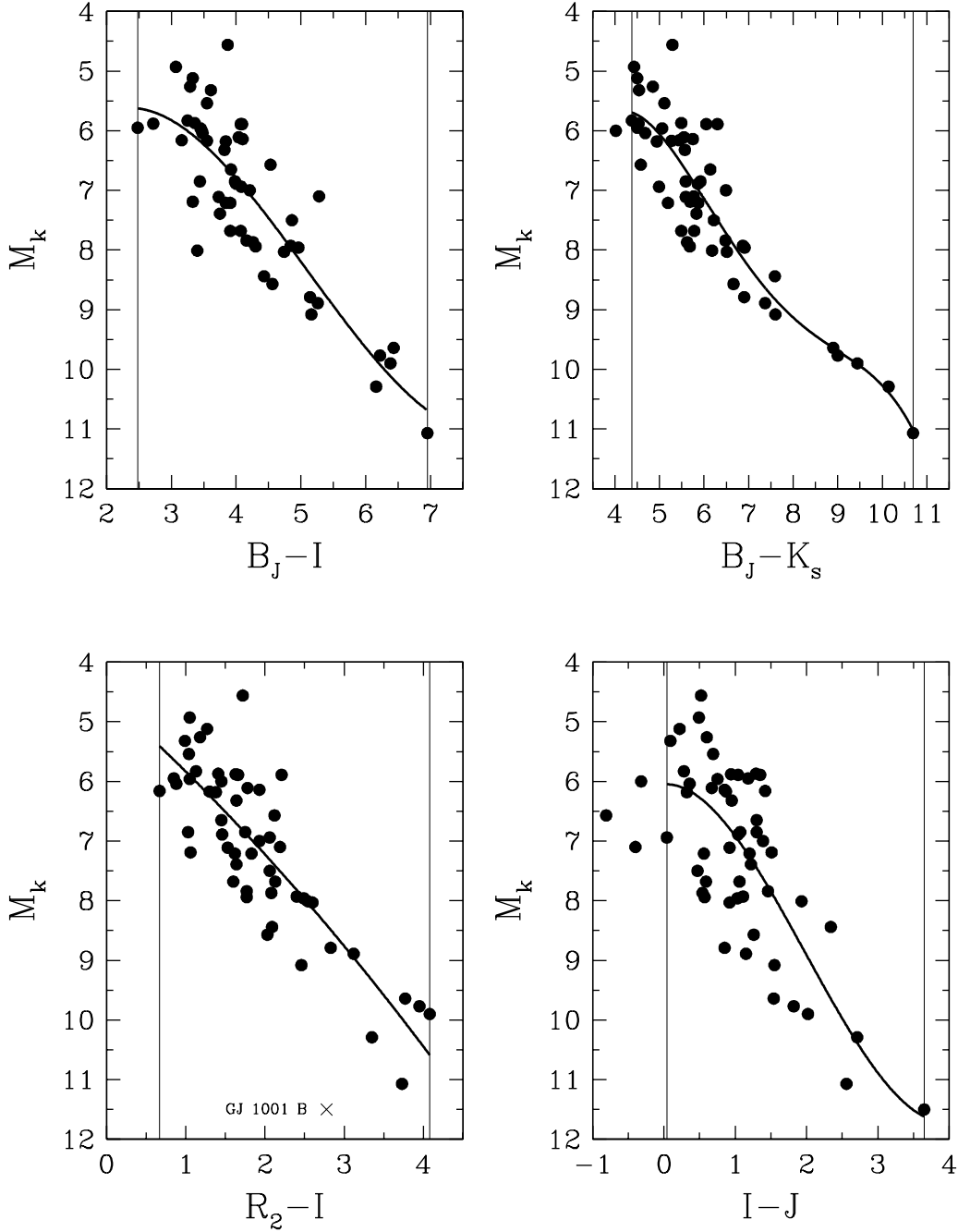}
\caption{Four representative $color-M_{Ks}$ relations used to estimate
distances to main sequence stars.  The points are for stars within 10
pc.  Details about the fits are given in Table~\ref{details}.  In
the $(R_2-I)$ plot, the X represents the L dwarf GJ 1001B $=$ LHS
102B, which was not used in the fit because the relation doubles back
at such faint $M_Ks$.}
\label{relations}
\end{figure*}

In total, there are 15 possible color-M$_{Ks}$ relations that can be
derived from the six bandpasses.  Of these, $(B_{\rm J}-R_2)$,
$(J-H)$, $(J-K)$, and $(H-K)$ are not useful because in each case the
range in color is quite restricted and does not predict reliable
M$_{Ks}$ values.  The remaining 11 relations are used as an ensemble
to generate up to 11 different distance estimates for each star,
provided that the star's color falls within the valid range.  Each
relation is locked to M$_{Ks}$ because practically all undiscovered
nearby stars (and many brown dwarfs) of interest will have a reliable
${Ks}$ in 2MASS.  Four exemplary relations are illustrated in 
Figure~\ref{relations}.
The $(B_{\rm J}-I)$ and $(B_{\rm J}-Ks)$ relations represent the
largest range in color for photographic only and photographic/infrared
colors.  The $(B_{\rm J}-Ks)$ relation, in particular, shows the great
strength in combining optical and infrared data, spanning more than
six full magnitudes in color.  The less reliable $(R_2-I)$ and $(I-J)$
relations are also illustrated.  In these cases, the colors span only
$\sim$3.5 magnitudes, and the scatter in the photographic magnitudes
compromises the relations.  Nonetheless, they still add weight to the
final ensemble distance estimates.  Details for each fit, including
the number of objects used to create the fit, its applicable range,
the coefficients, and the RMS in magnitudes, are given in
Table~\ref{details}.

To provide a measure of the ensemble technique's reliability, the
RECONS stars of known distance have been run back through the
relations.  Each star has up to 11 different distance estimates that
are combined to produce a mean distance estimate and error,
represented by the standard deviation of the individual estimates.
The average offset between the estimated and true distances is 26\%,
which is remarkable given the imprecise nature of the
photographic magnitudes, and of course, the intrinsic cosmic scatter in
the stars due to metallicity and age effects.  
These relations can be used within the stated ranges for single, main
sequence stars with reliable magnitudes found in SuperCOSMOS and 2MASS.  
The distances derived using these relations will, of course, not be
reliable for subdwarfs.  Even for main sequence stars, in a few cases the
effects of age and metallicity will be severe enough that the predicted
and true distances will differ by more than 50\%, as is the case for seven
of the 55 stars used in the relations.  In comparison, 14 of the 55 stars
have estimated distances within 10\% of their true distances.

\section {Discussion}

Table~\ref{distances} gives distance estimates and errors for the new
SCR discoveries and the previously known stars that were recovered
during the SCR search.  Two of the four red SCR stars are likely to be
within the 10 pc horizon of RECONS.  SCR1845--6357 is likely to be one
of the nearest few dozen stars.  Based on preliminary CCD $VRI$
photometry and photometric distance relations for white dwarfs, we
estimate a distance of $\sim$15--25 pc for SCR2012--5956.

Only 11 of the 23 previously known objects have trigonometric parallax
measurements, as given in the Yale Parallax Catalogue \citep{van95} and
from the Hipparcos mission \citep{esa97}.  In general those that have
trigonometric distances match the distance estimates well, although
LHS 128 and LHS 531 may be unresolved multiples because they are
significantly further than predicted.  All of the SCR discoveries and
previously known stars without trigonometric measurements that are
predicted to be within 25 pc have been included in our Cerro Tololo
Interamerican Observatory Parallax Investigation (CTIOPI), as
indicated in the notes column.  Several of these stars will likely
fall within the 10 pc horizon of the RECONS sample.

It is useful to assess briefly the complete, whole--sky sample of high
proper motion stars as it is currently known.  At Georgia State one of
the authors of this work (Jao) has compiled a comprehensive list of
objects with $\mu$ $>$ 1.0\arcsec/yr.  Although this list is the
subject of a future paper, we note here that the distribution of
systems is, as expected, tipped to the north.  As of January 1, 2003,
the counts of published systems in sky quartets of equal area are 143
from $\delta$ $=$ $+$90 to $+$30, 153 from $\delta$ $=$ $+$30 to
$+$00, 127 from $\delta$ $=$ $-$00 to $-$30, and 126 from $\delta$ $=$
$-$30 to $-$90, yielding a total of 549 systems.  The 8\%
overabundance in the north is an indicator that more fast moving
systems are likely to be found at southern declinations.  The five new
objects reported here are another step toward rectifying this
incompleteness in the southern sky.

\section {Conclusions}

The five new stars reported here provide important new nearby star
candidates, with proper motions ranking them in the top few hundred
stellar systems.  The two fastest movers, SCR1845--6357 and
SCR1138--7721, rank respectively as the 34th and 54th fastest proper
motion systems known.  These are merely the harbingers of a set of new
high proper motion objects that remain to be discovered in the
southern sky using SuperCOSMOS Sky Survey data.  In a follow-up paper
\citep{hen04} we present accurate optical photoelectric photometry
and spectroscopy of high proper motion objects in this portion of the
SCR survey, thereby revealing their true nature and allowing us to
refine the distance estimates.  In addition to the five stars
highlighted here, there are 116 additional new discoveries with $\mu$
$=$ 0.4--1.0\arcsec/yr that will also be the subjects of future
efforts.

Such nearby stars are, of course, useful for bolstering the population
statistics of the Galaxy, and move us closer to an accurate census of
the Sun's neighbors.  The momentum for discovering new nearby stars
comes from many directions, including the identification of systems
for stellar mass determinations, planet searches, the detection of
signatures of life, and SETI, simply because proximity is a key
element in any search in which resolution is required or the intrinsic
signals may be weak.  The SCR search reported here is merely in its
initial phase and will undoubtedly reveal many new nearby stars as we
continue to push northward.  This astrometric search is complementary
to photometric and spectroscopic searches for nearby stars, such as
\citet{rei02} and \citet{hen02}.  All of these searches continue to
yield new nearby stars, bringing about a sort of nearby star
renaissance, as large scale surveys and significant computer power are
turned to the discovery of solar neighbors.

\section {Acknowledgements}

Funding for the SuperCOSMOS Sky Survey is provided by the UK Particle Physics
and Astronomy Research Council. NCH would like to thank colleagues in the Wide
Field Astronomy Unit at Edinburgh for their work in making the SSS possible;
particular thanks got to Mike Read, Sue Tritton and Harvey MacGillivray.
Acknowledgements concerning the source photographic material of the SSS can
be found in the survey papers cited herein.

The RECONS team at Georgia State University wishes to thank NASA's
Space Interferometry Mission for its continued support of our study of
nearby stars.

This publication makes use of data products from the Two Micron All Sky
Survey, which is a joint project of the University of Massachusetts and
the Infrared Processing and Analysis Center/California Institute of
Technology, funded by the National Aeronautics and Space Administration
and the National Science Foundation.



\begin{deluxetable}{lcccrrccl}
\tablecaption{SuperCOSMOS--RECONS Search Data for Objects with $\mu$
$>$ 1.0\arcsec/yr and ${\rm R}<16.5$ found between $\delta$ $=$
$-57.5^\circ$ and $-90^\circ$.
\label{photography}}

\tablewidth{0pt}
\tabletypesize{\scriptsize}

\tablehead{\colhead{Object}&
           \colhead{RA\tablenotemark{a} ~~ (J2000.0) ~ Dec\tablenotemark{a}}&
           \colhead{$\mu$\tablenotemark{b}}&
           \colhead{$\sigma_\mu$\tablenotemark{b}}&
           \colhead{PA\tablenotemark{b}}&
           \colhead{$R_2$\tablenotemark{c}}&
           \colhead{($B_{\rm J}-R_2$)\tablenotemark{c}}&
           \colhead{($R_2-I$)\tablenotemark{c}}&
           \colhead{Notes}}
\startdata
\multicolumn{9}{l}{New discoveries:         }\\
SCR0342--6407  &  03 42 57.44 ~~ $-$64 07 56.4 & 1.071 & 0.023 & 141.42 &  15.13   &   2.04   &   2.79   & short time span for $\mu$        \\
SCR1138--7721  &  11 38 16.82 ~~ $-$77 21 48.8 & 2.141 & 0.007 & 286.77 &  14.12   &   2.33   &   2.66   & found twice                      \\
SCR1845--6357\tablenotemark{d}  &  18 45 05.09 ~~ $-$63 57 47.7 & 2.558 & 0.012 &  74.80 &  16.33   & \nodata  &   3.80   & blended in $B_{\rm J}$           \\
SCR1848--6855  &  18 48 21.14 ~~ $-$68 55 34.5 & 1.287 & 0.013 & 194.38 &  16.07   & \nodata  &   2.11   & blended in $B_{\rm J}$           \\
SCR2012--5956  &  20 12 31.79 ~~ $-$59 56 51.6 & 1.440 & 0.011 & 165.62 &  15.63   &   1.03   &   0.50   & blue object                      \\
\multicolumn{9}{l}{Previously known objects:}\\
LHS  124       &  00 49 29.05 ~~ $-$61 02 32.8 & 1.126 & 0.019 &  93.86 &  10.78   &   2.33   &   1.59   &                                  \\
LHS  128       &  00 57 19.78 ~~ $-$62 14 43.7 & 1.061 & 0.023 &  81.32 &   8.40   &   2.16   &   0.94   &                                  \\
LHS  145       &  01 43 00.99 ~~ $-$67 18 30.5 & 1.083 & 0.013 & 197.36 &  13.18   &   0.58   &   0.27   &                                  \\
LHS  150       &  02 07 23.25 ~~ $-$66 34 11.5 & 1.773 & 0.022 &  78.48 &   9.79   &   2.40   &   1.33   &                                  \\
LHS  160       &  02 52 22.18 ~~ $-$63 40 47.6 & 1.174 & 0.017 &  58.16 &   9.80   &   2.36   &   1.58   &                                  \\
LHS  195       &  04 38 22.35 ~~ $-$65 24 57.6 & 1.437 & 0.013 &  30.11 &   8.70   &   1.17   &   0.75   &                                  \\
LHS  199       &  04 55 57.72 ~~ $-$61 09 46.6 & 1.102 & 0.011 & 124.08 &  10.92   &   2.31   &   1.67   &                                  \\
LHS  204       &  05 13 05.30 ~~ $-$59 38 43.9 & 1.079 & 0.010 &  58.97 &   7.49   &   1.18   &   0.44   & poor photometry\tablenotemark{e} \\
LHS  205       &  05 16 59.72 ~~ $-$78 17 20.6 & 1.139 & 0.012 & 177.15 &  10.74   &   2.09   &   1.67   &                                  \\
LHS   34       &  07 53 08.15 ~~ $-$67 47 31.6 & 2.128 & 0.009 & 135.75 &  13.55   &   1.02   &   0.45   &                                  \\
LHS  263       &  09 17 05.36 ~~ $-$77 49 23.7 & 1.045 & 0.009 & 141.21 &  12.15   &   2.02   &   2.27   &                                  \\
LHS  268       &  09 24 20.94 ~~ $-$80 31 21.1 & 1.284 & 0.011 &  11.94 &   9.37   &   1.19   &   0.25   &                                  \\
LHS  271       &  09 42 46.45 ~~ $-$68 53 06.0 & 1.150 & 0.008 & 357.17 &  11.24   &   2.52   &   2.22   &                                  \\
LHS  328       &  12 28 40.09 ~~ $-$71 27 51.4 & 1.183 & 0.010 & 338.89 &  12.88   &   1.93   &   1.83   &                                  \\
LHS  329       &  12 28 43.10 ~~ $-$71 27 56.4 & 1.172 & 0.007 & 338.10 &  14.99   &   2.00   &   2.31   &                                  \\
LHS  475       &  19 20 54.36 ~~ $-$82 33 16.3 & 1.278 & 0.012 & 164.26 &  11.83   &   1.91   &   1.69   &                                  \\
LHS  493       &  20 28 03.78 ~~ $-$76 40 15.9 & 1.444 & 0.011 & 149.11 &  12.93   &   1.97   &   2.20   &                                  \\
LHS  499       &  20 51 41.64 ~~ $-$79 18 40.1 & 1.221 & 0.013 & 143.89 &  10.81   &   2.25   &   1.38   &                                  \\
PJH 4051       &  21 15 15.20 ~~ $-$75 41 52.4 & 1.079 & 0.009 & 143.58 &  13.37   &   2.10   &   2.24   & Pokorny et al.\ (2003)           \\
J2231--7515    &  22 30 33.46 ~~ $-$75 15 24.3 & 1.865 & 0.007 & 167.59 &  16.21   &   1.78   &   0.64   & Scholz et al.\ (2002)            \\
J2231--7514    &  22 30 39.95 ~~ $-$75 13 55.3 & 1.873 & 0.008 & 167.57 &  15.82   &   1.45   &   0.60   & Scholz et al.\ (2002)            \\
LHS  531       &  22 55 45.46 ~~ $-$75 27 31.4 & 1.484 & 0.011 & 224.20 &   9.39   &   2.17   &   1.93   &                                  \\
LHS  532       &  22 56 24.69 ~~ $-$60 03 49.4 & 1.076 & 0.011 & 208.73 &  13.19   &   2.21   &   2.53   &                                  \\
\enddata

\tablenotetext{a}{Coordinates are precessed and proper motion
corrected to equinox and epoch J2000.0, and are accurate to
$\pm0.3$~arcsec (eg.\ Hambly et al.\ 2001c).}

\tablenotetext{b}{Units of proper motion are arcsec/yr; position angle
(PA) is measured in degrees east of north.}

\tablenotetext{c}{As described in the text, individual passbands
magnitudes are only accurate to $\sim0.3^{\rm m}$, while color indices
are accurate to $\sim0.1^{\rm m}$. Passbands are photographic: $B_{\rm
J}$, $R_{\rm 59F}$, $I_{\rm IVN}$.}

\tablenotetext{d}{This object was also discovered in the final survey
of \citet{pok04}.}

\tablenotetext{e}{This object is very bright, and has a heavily
saturated $R_2$ image -- here, we have quoted $R_1$ (ESO--R)
photometry, but note that all the photographic photometry for this
object is potentially subject to large systematic errors.}

\end{deluxetable}



\begin{deluxetable}{lrrrcl}
\tablecaption{Infrared Photometry for Objects with
$\mu$ $>$ 1.0\arcsec/yr and ${\rm R}<16.5$ found between $\delta$ $=$
$-57.5^\circ$ and $-90^\circ$.
\label{photometry}}

\tablewidth{0pt}
\scriptsize

\tablehead{\colhead{Object}&
           \colhead{$J$}&
           \colhead{$H$}&
           \colhead{$Ks$}&
           \colhead{($R_2-Ks$)}&
           \colhead{Notes}}

\startdata
\multicolumn{6}{l}{New discoveries:}\\
SCR0342--6407 &  11.32  &  10.89  &  10.58  &   4.55   &              \\
SCR1138--7721 &   9.40  &   8.89  &   8.52  &   5.60   &              \\
SCR1845--6357 &   9.54  &   8.97  &   8.51  &   7.82   &  very red    \\
SCR1848--6855 &  11.89  &  11.40  &  11.10  &   4.97   &              \\
SCR2012--5956 &  14.93  &  15.23  &  15.41  &   0.22   &  white dwarf, $Ks$ suspect \\
\multicolumn{6}{l}{Previously known objects:}\\		     
LHS 124       &   8.63  &   8.09  &   7.84  &   2.94   &              \\
LHS 128       &   7.08  &   6.49  &   6.28  &   2.12   &              \\
LHS 145       &  12.87  &  12.66  &  12.58  &   0.60   &  white dwarf \\
LHS 150       &   8.13  &   7.61  &   7.36  &   2.42   &              \\
LHS 160       &   7.67  &   7.12  &   6.83  &   2.97   &              \\
LHS 195       &   8.51  &   8.19  &   8.09  &   0.61   &              \\
LHS 199       &   9.04  &   8.51  &   8.31  &   2.61   &              \\
LHS 204       &   8.32  &   8.05  &   8.00  &$-$0.51~~~&  poor photographic photometry \\
LHS 205       &   8.07  &   7.44  &   7.20  &   3.55   &              \\
LHS  34       &  12.73  &  12.48  &  12.36  &   1.19   &  white dwarf \\
LHS 263       &   8.33  &   7.77  &   7.45  &   4.71   &              \\
LHS 268       &   8.89  &   8.53  &   8.46  &   0.91   &              \\
LHS 271       &   7.95  &   7.39  &   7.04  &   4.20   &              \\
LHS 328       &   9.81  &   9.30  &   9.05  &   3.83   &              \\
LHS 329       &  10.98  &  10.50  &  10.18  &   4.81   &              \\
LHS 475       &   8.56  &   8.00  &   7.69  &   4.15   &              \\
LHS 493       &   9.36  &   8.88  &   8.60  &   4.33   &              \\
LHS 499       &   8.46  &   7.91  &   7.66  &   3.14   &              \\
PJH 4051      &  10.14  &   9.60  &   9.33  &   4.05   &              \\
J2231-7515    &  14.86  &  14.82  &  14.72  &   1.49   &  white dwarf \\
J2231-7514    &  14.66  &  14.66  &  14.44  &   1.38   &  white dwarf \\
LHS 531       &   6.62  &   6.08  &   5.81  &   3.58   &              \\
LHS 532       &   8.98  &   8.36  &   8.11  &   5.08   &              \\
	       	       		 	              
\enddata
\end{deluxetable}



\begin{deluxetable}{cccrrrrrc}

\tablecaption{Details of Photometric Distance Relations.
\label{details}}

\tablewidth{0pt}
\scriptsize

\tablehead{\colhead{}                    &
           \colhead{\# Stars}            &
           \colhead{Applicable}          &
           \colhead{Coeff \#1}           &
           \colhead{Coeff \#2}           &
           \colhead{Coeff \#3}           &
           \colhead{Coeff \#4}           &
           \colhead{Coeff \#5}           &
           \colhead{RMS}                 \\
           \colhead{Color}               &
           \colhead{in Fit}              &
           \colhead{Range}               &
           \colhead{$\times$ (color)$^4$}&
           \colhead{$\times$ (color)$^3$}&
           \colhead{$\times$ (color)$^2$}&
           \colhead{$\times$ (color)}    &
           \colhead{(constant)}          &
           \colhead{(mag)}}
\startdata
$(B_{\rm J}-R_2)$ & 54 &    not useful   &       \nodata  &      \nodata  &      \nodata  &      \nodata &      \nodata  &  \nodata \\
$(B_{\rm J}-I)$   & 54 &   2.48 to 6.95  &       \nodata  & $-$  0.06597  & $+$  1.00958  & $-$  3.65843 & $+$  9.49477  &  0.74  \\
$(B_{\rm J}-J)$   & 54 &   3.53 to 9.51  &  $+$  0.01720  & $-$  0.44789  & $+$  4.18392  & $-$ 15.61513 & $+$ 25.69047  &  0.62  \\
$(B_{\rm J}-H)$   & 54 &   4.15 to 10.22 &  $+$  0.01736  & $-$  0.49708  & $+$  5.13558  & $-$ 21.71069 & $+$ 37.74852  &  0.64  \\
$(B_{\rm J}-Ks)$  & 54 &   4.38 to 10.69 &  $+$  0.01385  & $-$  0.41706  & $+$  4.52981  & $-$ 20.08433 & $+$ 36.70961  &  0.63  \\
$(R_2-I)$         & 54 &   0.67 to 4.08  &       \nodata  &      \nodata  & $+$  0.07403  & $+$  1.16691 & $+$  4.59375  &  0.76  \\ 
$(R_2-J)$         & 55 &   1.08 to 6.43  &  $+$  0.03685  & $-$  0.53287  & $+$  2.68760  & $-$  4.56720 & $+$  8.21182  &  0.70  \\
$(R_2-H)$         & 55 &   1.68 to 7.49  &  $+$  0.02066  & $-$  0.37082  & $+$  2.36926  & $-$  5.37494 & $+$  9.74196  &  0.72  \\
$(R_2-Ks)$        & 55 &   1.92 to 8.15  &  $+$  0.01260  & $-$  0.25196  & $+$  1.78947  & $-$  4.36444 & $+$  9.10891  &  0.71  \\
$(I-J)$           & 55 &   0.04 to 3.65  &       \nodata  & $-$  0.19062  & $+$  1.13456  & $-$  0.07582 & $+$  6.05024  &  1.00  \\
$(I-H)$           & 55 &   0.61 to 4.71  &       \nodata  & $-$  0.17978  & $+$  1.40873  & $-$  1.58307 & $+$  6.60017  &  1.03  \\
$(I-Ks)$          & 55 &   0.91 to 5.37  &       \nodata  & $-$  0.16765  & $+$  1.47110  & $-$  2.23929 & $+$  7.04432  &  0.99  \\
$(J-H)$           & 55 &    not useful   &       \nodata  &      \nodata  &      \nodata  &      \nodata &      \nodata  &  \nodata \\
$(J-Ks)$          & 55 &    not useful   &       \nodata  &      \nodata  &      \nodata  &      \nodata &      \nodata  &  \nodata \\
$(H-Ks)$          & 55 &    not useful   &       \nodata  &      \nodata  &      \nodata  &      \nodata &      \nodata  &  \nodata \\

\enddata
\end{deluxetable}



\begin{deluxetable}{lccccl}

\tablecaption{Distance Estimates and True Distances for Objects with
$\mu$ $>$ 1.0\arcsec/yr and ${\rm R}<16.5$ found between $\delta$ $=$
$-57.5^\circ$ and $-90^\circ$.
\label{distances}}

\tablewidth{0pt}
\scriptsize

\tablehead{\colhead{Object}&
           \colhead{\# est}&
           \colhead{dist est}&
           \colhead{dist YPC}&
           \colhead{dist HIP}&
           \colhead{Notes}}

\startdata
\multicolumn{6}{l}{New discoveries:}\\
SCR0342--6407 & 11 &  39.3 $\pm$ 11.7 &      \nodata     &      \nodata     &  CTIOPI                                        \\
SCR1138--7721 & 11 &   8.8 $\pm$  1.7 &      \nodata     &      \nodata     &  CTIOPI                                        \\
SCR1845--6357 &  6 &   3.5 $\pm$  0.7 &      \nodata     &      \nodata     &  CTIOPI, no $B_{\rm J}$ data, $R_2-J$ too red  \\
SCR1848--6855 &  7 &  34.8 $\pm$  9.8 &      \nodata     &      \nodata     &  CTIOPI                                        \\
SCR2012--5956 &  0 &      \nodata     &      \nodata     &      \nodata     &  CTIOPI, white dwarf, dist 15--25 pc           \\
\multicolumn{6}{l}{Previously known objects:}\\
LHS 124       & 11 &  19.6 $\pm$  1.2 &      \nodata     &      \nodata     &  CTIOPI                                        \\                 
LHS 128       &  8 &  11.8 $\pm$  1.1 &  16.9 $\pm$  2.8 &  19.4 $\pm$  0.4 &  too blue in 3 colors, multiple?               \\
LHS 145       &  0 &      \nodata     &      \nodata     &      \nodata     &  CTIOPI, white dwarf                           \\
LHS 150       & 11 &  18.3 $\pm$  1.6 &  15.5 $\pm$  4.6 &      \nodata     &  CTIOPI                                        \\
LHS 160       & 11 &  12.2 $\pm$  0.8 &  12.3 $\pm$  2.4 &  11.5 $\pm$  0.3 &                                                \\
LHS 195       &  0 &      \nodata     &     not useful   &  58.8 $\pm$  3.4 &  too blue, too bright                          \\
LHS 199       & 11 &  26.9 $\pm$  3.2 &  22.5 $\pm$  5.8 &      \nodata     &  CTIOPI                                        \\
LHS 204       &  0 &      \nodata     &  35.1 $\pm$ 13.5 &  68.7 $\pm$  4.8 &  too blue, too bright                          \\
LHS 205       & 11 &  12.0 $\pm$  0.8 &  12.9 $\pm$  1.9 &      \nodata     &  CTIOPI                                        \\
LHS  34       &  0 &      \nodata     &   7.1 $\pm$  0.4 &      \nodata     &  CTIOPI, white dwarf                           \\
LHS 263       & 11 &   8.2 $\pm$  1.2 &      \nodata     &      \nodata     &  CTIOPI                                        \\
LHS 268       &  0 &      \nodata     &  46.3 $\pm$ 21.2 &  60.8 $\pm$  3.7 &  too blue, too bright                          \\
LHS 271       & 11 &   8.0 $\pm$  1.4 &      \nodata     &      \nodata     &  CTIOPI                                        \\
LHS 328       & 11 &  25.1 $\pm$  3.1 &      \nodata     &      \nodata     &                                                \\
LHS 329       & 11 &  27.4 $\pm$  4.6 &      \nodata     &      \nodata     &                                                \\
LHS 475       & 11 &  11.8 $\pm$  3.0 &      \nodata     &      \nodata     &  CTIOPI                                        \\
LHS 493       & 11 &  16.5 $\pm$  1.9 &      \nodata     &      \nodata     &  CTIOPI                                        \\
LHS 499       & 11 &  16.7 $\pm$  1.3 &  15.9 $\pm$  3.1 &      \nodata     &                                                \\
PJH 4051      & 11 &  26.1 $\pm$  3.5 &      \nodata     &      \nodata     &                                                \\
J2231-7515    &  0 &      \nodata     &      \nodata     &      \nodata     &  CTIOPI, white dwarf                           \\
J2231-7514    &  0 &      \nodata     &      \nodata     &      \nodata     &  CTIOPI, white dwarf                           \\
LHS 531       & 11 &   6.2 $\pm$  0.5 &   8.3 $\pm$  0.7 &   8.6 $\pm$  0.1 &  multiple?                                     \\
LHS 532       & 11 &   9.1 $\pm$  1.2 &      \nodata     &      \nodata     &  CTIOPI                                        \\

\enddata
\end{deluxetable}


\end{document}